# Identification of the Steering and Speed Systems of a Four-Wheel-Steering Tractor for Optimal Control

**Riikka Soitinaho** * **Timo Oksanen** *

* *Technical University of Munich, Germany; Chair of Agrimechatronics; Munich Institute of Robotics and Machine Intelligence (MIRMI), (e-mail: firstname.surname@tum.de)*

**Abstract:** Model-based control design requires sufficiently accurate model of the system-to-be-controlled. This paper addresses the system identification of steering and speed control of a four-wheel-steering agricultural tractor for the purpose of developing path tracking control. To this end, we investigate the steering and speed control systems of the tractor. These are complex systems with digital, mechatronic, and hydraulic components challenging to model based on first principles. We take a data-driven approach to estimating the system models. The resulting model combines the kinematic model of the vehicle, actuators modelled as first-order systems, and estimated values for time constants and transport delays.

## 1. INTRODUCTION

Autonomous operation of the vehicles involved in agricultural production can help with production efficiency, labour shortages, challenging driving conditions, and reducing operator fatigue. Advanced model-based control systems are increasingly investigated and applied in autonomous driving. A sufficiently accurate system model is especially essential for model-based control.

Our aim is to control a four-wheel-steering (4WS) agricultural tractor to drive autonomously on the field. For this purpose, we investigate the steering and speed control systems of the tractor.

The power steering system of the tractor includes hydraulic components that are controlled with an electrohydraulic valve. The steering dynamics therefore differ from those of passenger cars, for example.

We aim to capture the dynamics of the steering and speed systems themselves, i.e. from steering angle setpoint (u) to the actual steering angle (y). We will use the resulting models in our control design to consider the steering and speed response to the control input. In this paper we focus on the system identification, not the control design.

Actuating the tractor to drive autonomously relies on complex systems with a variety of digital, mechatronic, and hydraulic components. Modeling these systems would be challenging based on first principles, or such a model would be unnecessarily complex. In system identification, a model of the system is determined based on observed input-output data. The estimation involves finding a suitable model set and the model parameters in that set. While this allows constructing a model without physical modeling, other issues need to be considered. The model should explain the estimation data "sufficiently" well, but not be too complex, in order to not overfit to the estimation data. It is beneficial to use insight into the nature of the system in order to reduce the model set Ljung (2010). Therefore, we take a data-driven gray-box approach to system identification of the steering and speed control systems.

⋆ This research was funded by the Deutsche Forschungsgemeinschaft (DFG, German Research Foundation) - 528103308.

## 2. BACKGROUND

System identification is a methodology to building mathematical models based on experimental data. Data-driven modeling, i.e. contructing models based on observed input-output data, is a fundamental technique in scientific modeling, and it has applications in various fields. Oftentimes the specific values of parameters are to be found within a chosen model structure, referred to as parameter estimation.

System modeling is important in mobile robotics for both model based control methods and to produce accurate models for simulations. Modeling a variety of different kinematics and modes of actuation has been previously explored. In indoor conditions, a typical choice of movement is based on differential drive or omnidirectional wheels.

Modeling such systems has been studied for example by Ardiani et al. (2021), who compare least-squares and instrumental variables techniques for parameter estimation of a differential drive mobile robot. Eschmann et al. (2021), on the other hand, use Gaussian process regression to estimate a data-driven model for an omnidirectional robot. The system inputs are the translational velocities in the x- and y-directions in the local robot coordinate frame, and the angular velocity about the z-axis. The model is used to predict the state of the mobile robot, i.e. its x- and y-coordinates and orientation.

However, a typical vehicle used in outdoor conditions is a nonholonomic vehicle often steered with the front wheels. Rödönyi et al. (2021) identify the lateral dynamics of a passenger car with a black-box approach based on a neural network based subspace-encoder method. The system inputs are the vehicle speed and the control signal to the steering, and the system output is the yaw-rate.

The question of whether data-driven models are sufficient to predict the behaviour of real-world vehicle systems has been studied e.g. by Vicente et al. (2021), who compare data-driven and physical modeling of lateral-longitudinal vehicle dynamics for a front-wheel steered vehicle. For the data-driven approach they estimate a linear time-invariant state-space model with a subspace identification technique. They conclude that the linear data-driven model outperformed both linear and nonlinear physical models in the real-world driving dataset of their study.

Vehicle system identification is often motivated by the need to establish a system model for model based control. Soitinaho and Oksanen (2023) investigate nonlinear model predictive control (MPC) based path tracking that utilizes a system model based on kinematics and system identification of the steering and speed actuation. The model is successfully applied in the MPC path tracking in field experiments. The vehicle is only front-wheel steered.

In this paper we focus specifically on a 4WS standard agricultural tractor. We present an analysis of the system under investigation and our insight on the modeling. We specifically investigate a 4WS standard tractor with independently controlled front- and rear-wheel steering, enabling also so-called crab steering. We give a detailed account of the experiment design, and present the system identification on the steering and speed systems. Our aim is to utilize the resulting model in model predictive control based path tracking for a 4WS agricultural tractor, and we want to develop our system model based on data collected in field conditions. A sufficent fit to the validation data suggests that the model would be feasible for field experiments in a model predictive path tracking controller.

## 3. SYSTEM DESCRIPTION

The research tractor we investigate is a modified Lindner Lintrac 130 (Traktorenwerk Lindner, Kundl, Austria) we call "G-trac" (see Fig. 1). G-trac is a 4WS agricultural tractor with a diesel engine, power of 100 kW (136 hp), a power-split CVT transmission, a wheelbase of 2.42 meters, track width of 1.70 meters, and an empty weight of 4520 kg. The maximum steering angle for the front and rear axle are 34 degrees and 15 degrees, respectively. Four-wheel-steering allows for both heading and course to be controlled.

The base in-vehicle-communication of G-trac uses J1939 CAN bus. G-trac is equipped with various sensor systems, and the measured data for position, heading, wheel speed, and steering angles is available over CAN bus. The control interface to autonomous driving functions including steering and speed control also works over CAN bus. The unit

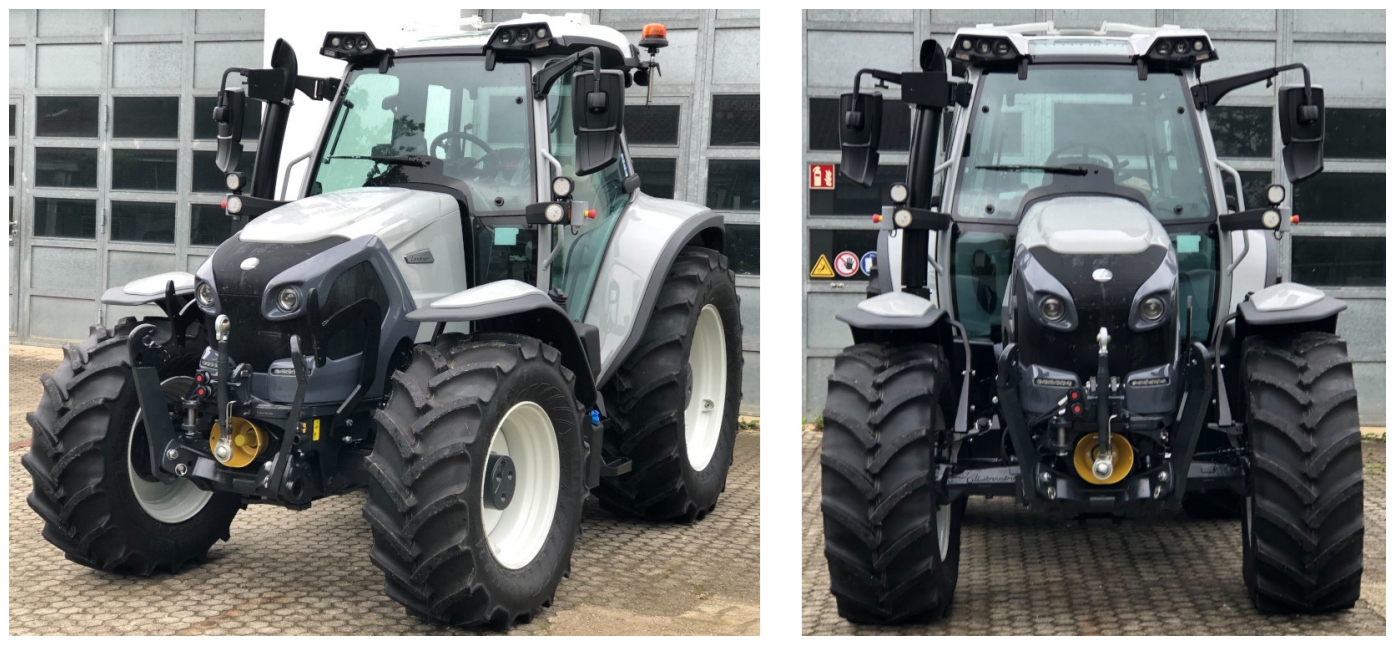

Fig. 1. Agricultural research tractor G-trac, a modified Lindner Lintrac 130.

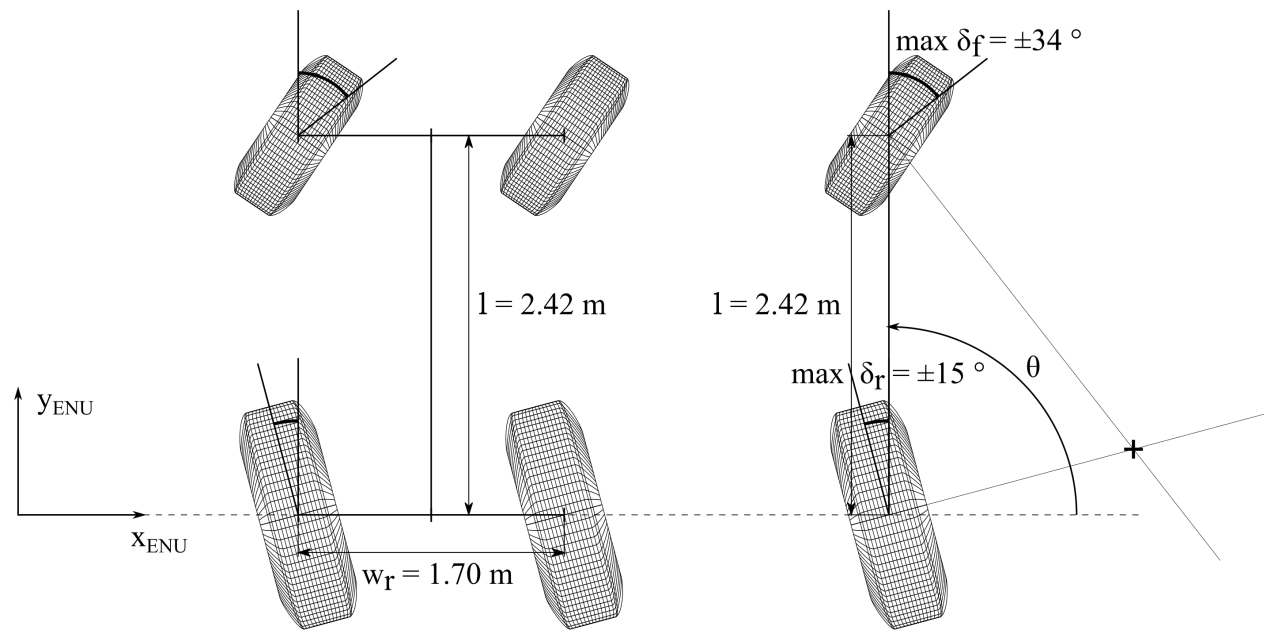

Fig. 2. Bicycle model of the 4WS research tractor G-trac.

for steering angles is degrees, and meters per second for speed.

For the control of the front-wheel steering we use the built-in Danfoss system with an electro-hydraulic steering valve. The rear-wheel steering system was redeveloped utilizing the original valves and a Parker IQAN mobile control system (see Hefele et al. (2024) for details).

G-trac has an autonomous mode and a manual mode. In the autonomous mode we control G-trac with Simulink Real-Time using Speedgoat rapid control prototyping (RCP) hardware, while the control design is developed in Matlab/Simulink.

### 3.1 Kinematic model

The basis of our system model for G-trac is the kinematic bicycle model of a 4WS vehicle (see Fig. 2). The reference point, the origin of the vehicle coordinate frame, is the center of the rear axle, and the vehicle state $\mathbf{X}$ is described by

$$\mathbf{X} = [x\ y\ \theta\ \delta_f\ \delta_r\ v]^T, \tag{1}$$

where $x$ and $y$ are the coordinates of the origin, $\theta$ is the heading, $\delta_f$ and $\delta_r$ are the front and rear wheel steering angles, and $v$ is the magnitude of the current velocity. The kinematics are described by the differential equation (2), where $l$ is the wheelbase of the vehicle.

$$\dot{\mathbf{X}} = \begin{bmatrix} \dot{x} \\ \dot{y} \\ \dot{\theta} \\ \dot{\delta}_f \\ \dot{\delta}_r \\ \dot{v} \end{bmatrix} = \begin{bmatrix} v \cdot \cos(\theta + \delta_r) \\ v \cdot \sin(\theta + \delta_r) \\ (v \cdot \sin(\delta_f - \delta_r)/(l \cdot \cos(\delta_f))) \\ \omega_f \\ \omega_r \\ a \end{bmatrix} \quad (2)$$

### *3.2 Steering actuation system*

G-trac has two independently controlled steering systems, front wheel steering and rear wheel steering.

We model both steering actuator responses as first-order systems with single input and single output, i.e. a transfer function with one pole as given by (3) and (4):

$$G_{\delta_f}(s) = \frac{K_{\delta_f}}{\tau_{\delta_f} s + 1} \cdot e^{-T_{\delta_f} s} \quad (3)$$

$$G_{\delta_r}(s) = \frac{K_{\delta_r}}{\tau_{\delta_r} s + 1} \cdot e^{-T_{\delta_r} s} \quad (4)$$

where $\tau_{\delta_f}$ and $\tau_{\delta_r}$ are the time constants of the actuator dynamics, $K_{\delta_f}$ and $K_{\delta_r}$ are the steady state gains, and $T_{\delta_f}$ and $T_{\delta_r}$ are the transport delays. The steady state gains for both actuator systems are assumed to be 1, i.e. $K_{\delta_f} = 1$ and $K_{\delta_r} = 1$. Steady state gain of 1 means that the actuator is able to eventually obtain the desired control value. With this assumption the steady state gains are set as known for the system identification. Therefore, the steering systems' identification consists of estimating the parameters $\tau_{\delta_f}$, $\tau_{\delta_r}$, $T_{\delta_f}$, and $T_{\delta_r}$. Previously e.g. Backman et al. (2012) have successfully applied a similar assumption about first order dynamics.

### *3.3 Speed actuation system*

Similar to the steering systems, we model the speed actuator response with a transfer function with one pole as given by (5):

$$G_v(s) = \frac{K_v}{\tau_v s + 1} \cdot e^{-T_v s} \quad (5)$$

where $\tau_v$ is the time constant of the actuator dynamics, $K_v$ is the steady state gain, and $T_v$ is the transport delay. The steady state gain is again assumed to be 1, i.e. $K_v = 1$. Therefore, the speed system identification consists of estimating the parameters $\tau_v$ and $T_v$.

### *3.4 Control system software*

We control G-trac with Simulink Real-Time using the RCP hardware Speedgoat. The experiment design for the system identification is implemented on a development computer using Matlab/Simulink, and the real-time application is built and downloaded from Simulink to the Speedgoat target computer, which runs a Simulink Real-Time kernel. The target computer is Speedgoat Baseline Real-Time Target Machine with Intel Celeron 2 GHz 4 cores and 4GB DDR3 RAM memory.

## 4. EXPERIMENTAL PROCEDURE

The experiments to record the estimation data were performed in the autumn after the growing season on an agricultural field in Southern Germany, on an area of the field where the maximum difference in elevation between the highest and lowest point is approximately 1.6 meters. The field had been planted in the spring with a mix of grass, clover, and alfalfa, and mowed during the summer. The height of the crop was approximately 10 centimeters at the time of the experiments. The weather on the day of the experiments was cloudy but no rain. All the experiments were performed on the same day.

### *4.1 Design of key experiments*

In the autonomous mode the steering and speed of G-trac are actuated according to the control input signals which are sent over CAN bus. Several experiments were performed with different excitation signals on each of the three actuation systems. Each experiment was started and stopped manually by the operator, while the rest of the experiment (i.e. sending the excitation signal, tractor actuation, and data recording) was automated.

In the experiments we recorded the actuator system response to several different square wave excitation signals to gather input-output data for the system identification. The square waves elicit a series of different step responses from the actuator systems.

During all experiments the engine of G-trac was set to run at a constant 1430 RPM. The excitation signal was sent to the actuator under testing, while the other two actuators received constant value control signals. For recording input-output data of the steering actuation, G-trac was set to drive at a constant speed, and for recording input-output data of the speed actuation, G-trac was set to steer straight ahead (zero degree steering angle), i.e.:

- Front steering $u_{\delta_f} \rightarrow$ rear steering 0.0°, speed 1.0 $\frac{\text{m}}{\text{s}}$
- Rear steering $u_{\delta_r} \rightarrow$ front steering 0.0°; speed 1.0 $\frac{\text{m}}{\text{s}}$
- Speed $u_v \rightarrow$ front steering 0.0°; rear steering 0.0°

### *4.2 Excitation signals and data recording*

The set of excitation data is described in table 1 for all three actuation systems. Column "Id" identifies data from individual experiments, and column "Description" defines the excitation signal to the actuation system under investigation. Last two columns give the constant value control signal that was sent to the other two actuators.

For the steering excitation signals we used symmetrical bipolar square waves with a 10 s phase width ($T$), a 50 % duty cycle, and different peak amplitudes ($A$) of up to the maximum steering angle both left and right. In addition, the excitation signals included square waves with interphase ($d_1$) and interpulse delay ($d_2$) of 10 s (i.e. return to zero degree for 10 s between positive and negative phase of the waveform). For speed actuation excitation signals we used unipolar square waves with a 10 s phase width, a 50 % duty cycle, and a peak amplitude of up to 2.34 $\frac{\text{m}}{\text{s}}$, however with a slew rate of 0.84 $\frac{\text{m}}{\text{s}^2}$, which was

Table 1. Description of all recorded excitation-response data for all three actuator systems.

**Front wheel steering**

| Id | Description of $u_{\delta_f}$ | $u_{\delta_r}\,(^\circ)$ | $u_v\,(\frac{\mathrm{m}}{\mathrm{s}})$ |
|---|---|---|---|
| 1 | $A = \pm\,5^\circ$, $T = 10\,\mathrm{s}$, duty cycle 50 % | 0° | 1.0 |
| 2 | $A = \pm\,10^\circ$, $T = 10\,\mathrm{s}$, duty cycle 50 % | 0° | 1.0 |
| 3 | $A = \pm\,20^\circ$, $T = 10\,\mathrm{s}$, duty cycle 50 % | 0° | 1.0 |
| 4* | $A = \pm\,30^\circ$, $T = 10\,\mathrm{s}$, duty cycle 50 % | 0° | 1.0 |
| 5 | $A = \pm\,34^\circ$, $T = 10\,\mathrm{s}$, duty cycle 50 % | 0° | 1.0 |
| 6 | $A = \pm\,10^\circ$, $T = 10\,\mathrm{s}$, $d_1 = 10\,\mathrm{s}$, $d_2 = 10\,\mathrm{s}$ | 0° | 1.0 |
| 7 | $A = \pm\,34^\circ$, $T = 10\,\mathrm{s}$, $d_1 = 10\,\mathrm{s}$, $d_2 = 10\,\mathrm{s}$ | 0° | 1.0 |

**Rear wheel steering**

| Id | Description of $u_{\delta_r}$ | $u_{\delta_f}\,(^\circ)$ | $u_v\,(\frac{\mathrm{m}}{\mathrm{s}})$ |
|---|---|---|---|
| 1 | $A = \pm\,5^\circ$, $T = 20\,\mathrm{s}$, duty cycle 50 % | 0° | 1.0 |
| 2 | $A = \pm\,10^\circ$, $T = 20\,\mathrm{s}$, duty cycle 50 % | 0° | 1.0 |
| 3* | $A = \pm\,15^\circ$, $T = 20\,\mathrm{s}$, duty cycle 50 % | 0° | 1.0 |
| 4 | $A = \pm\,5^\circ$, $T = 20\,\mathrm{s}$, $d_1 = 10\,\mathrm{s}$, $d_2 = 10\,\mathrm{s}$ | 0° | 1.0 |
| 5 | $A = \pm\,15^\circ$, $T = 20\,\mathrm{s}$, $d_1 = 10\,\mathrm{s}$, $d_2 = 10\,\mathrm{s}$ | 0° | 1.0 |

**Speed**

| Id | Description of $u_v$ | $u_{\delta_f}\,(^\circ)$ | $u_{\delta_r}\,(^\circ)$ |
|---|---|---|---|
| 1 | $A = 0.83\,\frac{\mathrm{m}}{\mathrm{s}}$, $T = 20\,\mathrm{s}$, duty cycle 50 % | 0° | 0° |
| 2 | $A = 1.55\,\frac{\mathrm{m}}{\mathrm{s}}$, $T = 20\,\mathrm{s}$, duty cycle 50 % | 0° | 0° |
| 3* | $A = 2.34\,\frac{\mathrm{m}}{\mathrm{s}}$, $T = 20\,\mathrm{s}$, duty cycle 50 % | 0° | 0° |

*This input-output data was used for model estimation.

earlier estimated from the steering actuator response. Each experiment (amplitude, waveform, actuation system) was performed a single time.

An example of the excitation-response data is visualized in Fig. 3 for the front-wheel steering, Fig. 4 for the rear-wheel steering, and Fig. 5 for the speed actuation. Dashed line shows the excitation (reference) signal and solid green line shows the response based on the sensor measuring the actuator state. For rear-wheel steering and speed actuation these are the data that was used for the parameter estimation.

Any transient in the actuator response that is due to the misalignment between the initial actuator position and the reference value in the beginning of the excitation signal was removed from the data. A combination of the excitation-response data from the experiments was used for the system identification and validation of the identified system models.

### *4.3 System identification method*

In the actuator system identification we estimate a process model for a single-input single-output (SISO) system based on time-domain data. We choose to model the actuator systems with a first-order system transfer function, which has a steady-state gain $K$, time constant $\tau$, and transport delay $T$. The same model structure is used for all actuator systems (front-wheel steering, rear-wheel steering, and speed actuation). A unit steady-state gain is assumed based on the assumption that the actuator is eventually able to reach the reference value. Therefore, $K = 1$ is set as known for estimation of all system models.

The system identification is performed using the MATLAB system identification toolbox that performs iterative parameter estimation, iterative meaning that the estimates of the unknown model parameters are repeatedly refined by minimizing the error between response data from the real system and the model prediction. Adaptive subspace Gauss-Newton search is used as the numerical search method for the iterative parameter estimation.

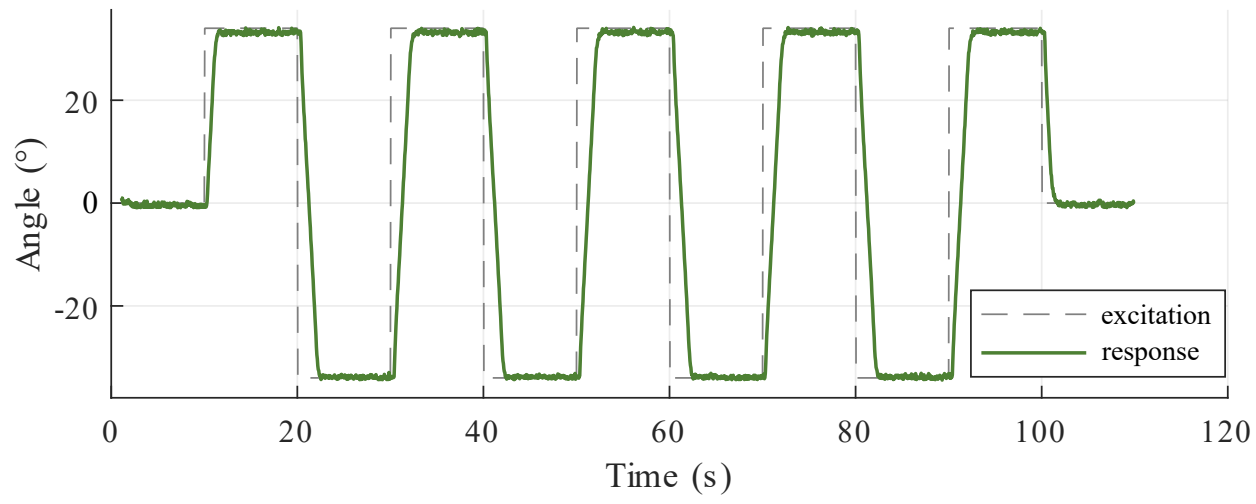

Fig. 3. Front steering estimation data Id 5.

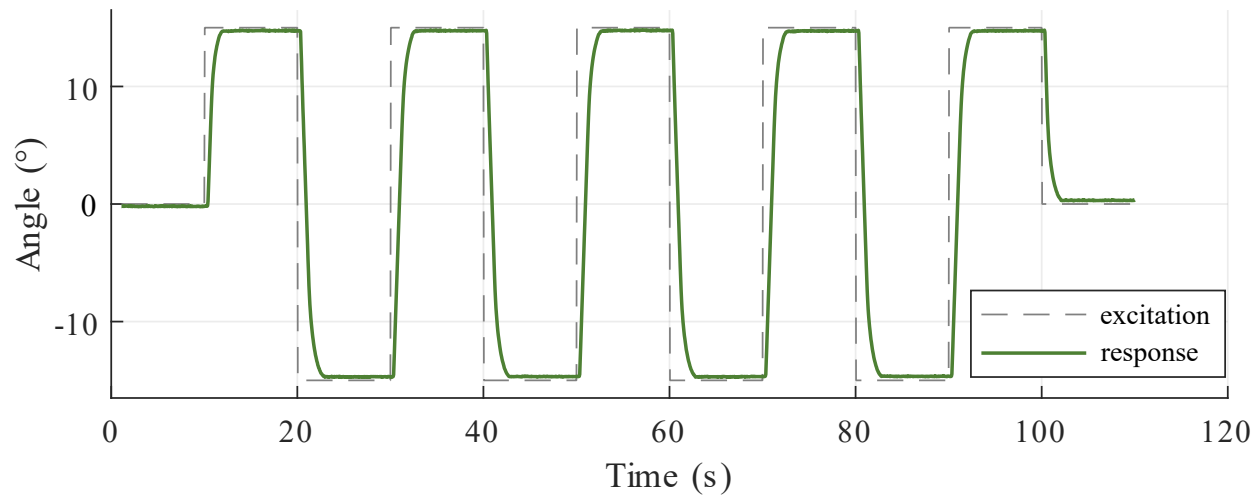

Fig. 4. Rear steering estimation data Id 3.

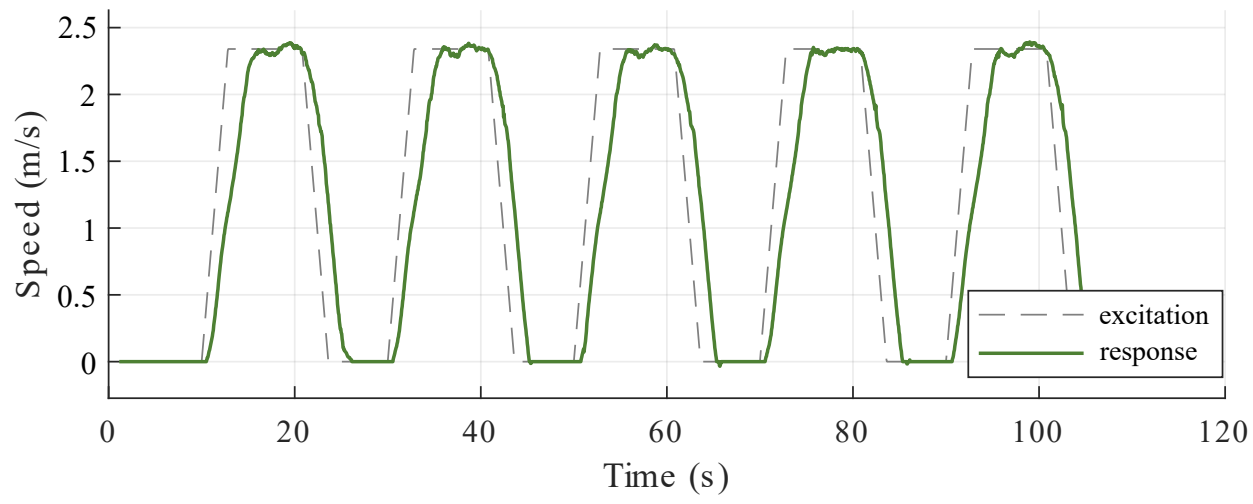

Fig. 5. Speed actuation estimation data Id 3.

## 5. RESULTS

The parameters values that were found in the system identification for each actuation system are:

- Front steering: $\tau_{\delta_f} = 0.73$, $T_{\delta_f} = 0.43$, and $K_{\delta_f} = 1$ ($\mathrm{sd}(\tau_{\delta_f}) = 0.0066$, $\mathrm{sd}(T_{\delta_f}) = 0.0046$, and fit to estimation data 91.5 %)
- Rear steering: $\tau_{\delta_r} = 0.62$, $T_{\delta_r} = 0.39$, and $K_{\delta_r} = 1$ ($\mathrm{sd}(\tau_{\delta_r}) = 0.0035$, $\mathrm{sd}(T_{\delta_r}) = 0.0024$, and fit to estimation data 95.1 %)
- Speed: $\tau_v = 1.13$, $T_v = 0.39$, and $K_v = 1$ ($\mathrm{sd}(\tau_v) = 0.0135$, $\mathrm{sd}(T_v) = 0.0117$, and fit to estimation data 91.0 %)

Table 2 shows the fit of simulated model output to the real system response for each of the recorded estimation data. Simulated model output means the simulated response of the estimated model to the corresponding excitation signal.

For the estimated speed actuation model, the fit to the estimation data was 91.0 %, and the fit to validation data was between 84.9 % and 93.2 %. For the front wheel

Table 2. Fit of simulated output of estimated model to input-output data from real system.

**Front wheel steering**

| Id | 1 | 2 | 3 | 4* | 5 | 6 | 7 |
|---|---|---|---|---|---|---|---|
| Fit (%) | 66.5 | 73.8 | 84.0 | 91.6 | 89.4 | 76.4 | 87.3 |

**Rear wheel steering**

| Id | 1 | 2 | 3* | 4 | 5 |
|---|---|---|---|---|---|
| Fit (%) | 85.7 | 91.4 | 95.1 | 89.1 | 90.0 |

**Speed**

| Id | 1 | 2 | 3* |
|---|---|---|---|
| Fit (%) | 84.9 | 93.2 | 91.0 |

steering actuation model, the fit to the estimation data was 91.5 %, and the fit to validation data was between 66.5 % and 89.4 %. For the rear wheel steering actuation model, the fit to the estimation data was 95.1 %, and the fit to validation data was between 85.7 % and 91.4 %.

The identified model for the speed actuation has a time delay of 390 ms, front wheel steering actuation has a time delay of 430 ms, and rear wheel steering actuation a time delay of 390 ms.

Fig. 6 (front steering), Fig. 7 (rear steering), and Fig. 8 (speed actuation) show the simulated model output in comparison to the real system response to the corresponding excitation data for data from selected experiments of the dataset. Excitation signal is shown in dashed gray, real system response in solid green, and simulated model output from the estimated model in solid black. The corresponding simulated model output fit for Fig. 6 (data Id 7) is 87.3 %, Fig. 7 (data Id 2) is 91.4 %, and Fig. 8 (data Id 1) 84.9 %.

Finally, the median of the tracking error distribution for a maximum steering angle step input after 5 % settling time was calculated to assess the assumption on unit steady state gain. For rear steering the calculated median of -0.26°is 1.7 % below the control input signal of 15.0°. For front steering the calculated median of -0.15°is 0.4 % below the control input signal of 34.0°.

The final model for the 4WS vehicle system consists of the kinematic model and the equations for the identified actuator system models. The final system model is therefore given by (6)

$$\dot{\mathbf{x}} = \begin{bmatrix} \dot{x} \\ \dot{y} \\ \dot{\theta} \\ \dot{\delta_f} \\ \dot{\delta_r} \\ \dot{v} \end{bmatrix} = \begin{bmatrix} v \cdot \cos(\theta + \delta_r) \\ v \cdot \sin(\theta + \delta_r) \\ (v \cdot \sin(\delta_f - \delta_r)/(l \cdot \cos(\delta_f))) \\ -\frac{1}{\tau_{\delta_f}}\delta_f + \frac{K_{\delta_f}}{\tau_{\delta_f}} u_{\delta_f} \\ -\frac{1}{\tau_{\delta_r}}\delta_r + \frac{K_{\delta_r}}{\tau_{\delta_r}} u_{\delta_r} \\ -\frac{1}{\tau_v} v + \frac{K_v}{\tau_v} u_v \end{bmatrix} \quad (6)$$

where equations for $\dot{\delta_f}$, $\dot{\delta_r}$, and $\dot{v}$ are the ordinary differential equations that correspond to the transfer functions (3), (4), and (5), and the parameters take values $\tau_{\delta_f} = 0.73$, $T_{\delta_f} = 0.43$, $K_{\delta_f} = 1$, $\tau_{\delta_r} = 0.62$, $T_{\delta_r} = 0.39$, $K_{\delta_r} = 1$, $\tau_v = 1.13$, $T_v = 0.39$, and $K_v = 1$. Finally, the control

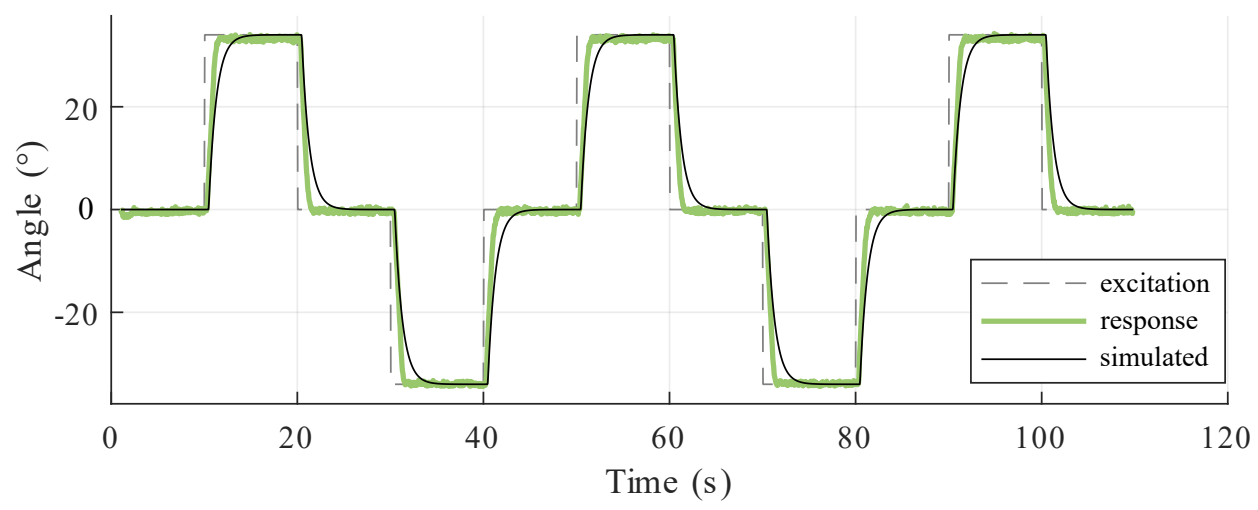

Fig. 6. Simulated model output front steering data Id 7.

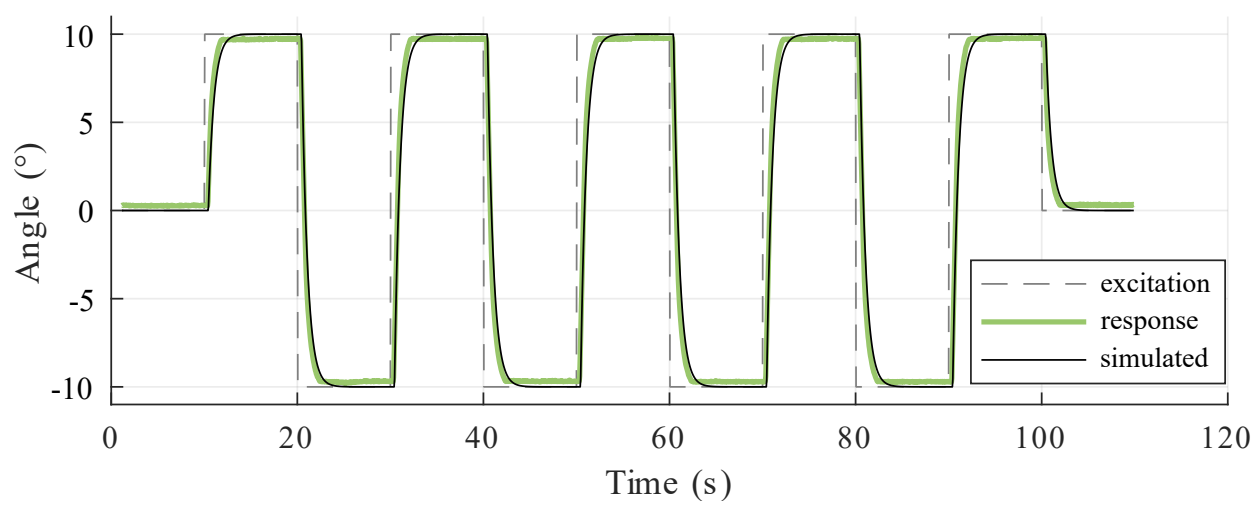

Fig. 7. Simulated model output rear steering data Id 2.

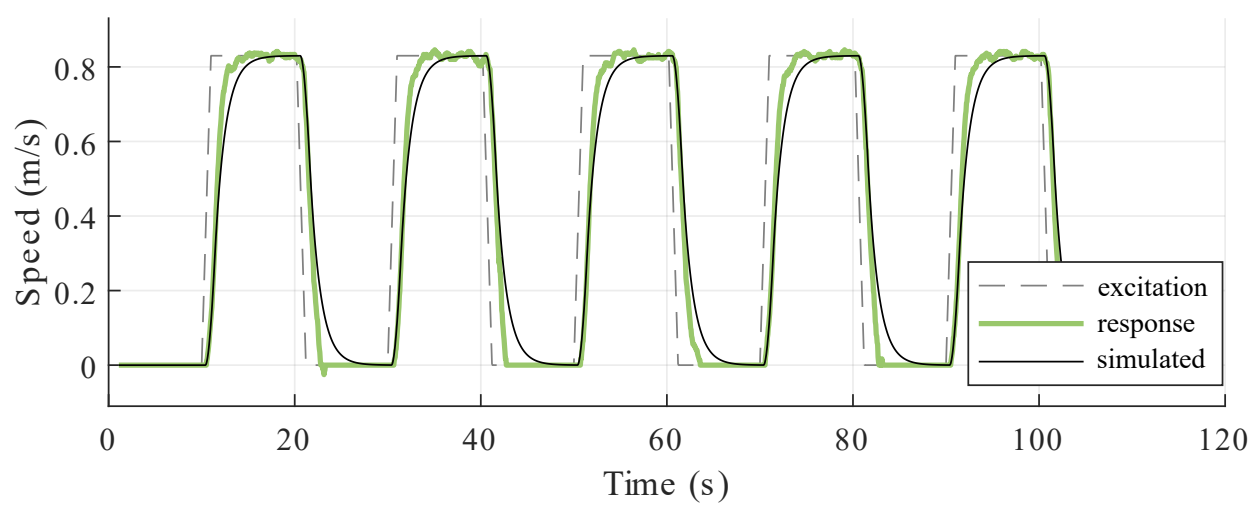

Fig. 8. Simulated model output speed actuation data Id 1.

input vector of the system is

$$\mathbf{u} = \begin{bmatrix} u_{\delta_f} & u_{\delta_r} & u_v \end{bmatrix}^T \quad (7)$$

where $u_{\delta_f}$, $u_{\delta_r}$, and $u_v$ are the inputs for front wheel steering, rear wheel steering, and speed, respectively.

## 6. DISCUSSION

This paper proposed a system identification procedure and a resulting system model for a standard 4WS agricultural tractor.

A dataset of several experiments that were recorded in field conditions was collected for estimation and validation of the system model of the speed actuation system, and the front- and rear-wheel steering systems independently.

While the dataset covers the relevant operating range of the actuators, an added variety of excitation signals could be applied to further asses the model response. For example, a typical turn manoeuvre from a path tracking controller. A sensitivity analysis of the identified model parameters is planned for the future.

The system model is intended to be applied in optimal control based path tracking for a 4WS tractor. It specifically models the research tractor G-trac, however, the system identification procedure can be adapted to model similar 4WS or front-wheel-steering vehicles. Such models are needed for example in model based control systems that apply model predictive control.

A limitation of the model is that it does not consider slip in its current formulation. This means that the model has limited applicability to heavy work such as ploughing. Our immediate use-case however is grasslands operations like mowing. A system model that explicitly addresses slip, as well as slip estimation, is left for future work.

## 7. CONCLUSION

This paper proposed a system model for a 4WS agricultural tractor for the purpose of optimal control based path tracking in low slip conditions. A system model of the vehicle was constructed based on the kinematic bicycle model of a 4WS vehicle and a data-driven approach for estimating the parameters of first-order systems with unit gain to model the steering and speed actuation. A sufficent fit to the validation data suggests that the model would be feasible for field experiments.

## ACKNOWLEDGEMENTS

We would like to thank Traktorenwerk Lindner GmbH.